# Control of Tension-Compression Asymmetry in Ogden Hyperelasticity with Application to Soft Tissue Modelling

*Kevin M. Moerman[a,b]\*, Ciaran K. Simms[b], Thomas Nagel[c]*

a) Massachusetts Institute of Technology, Media-Lab., Center for Extreme Bionics, Cambridge, MA, USA
b) Trinity Centre for Bioengineering, School of Engineering, Parsons Building, Trinity College, Dublin 2, Ireland
c) Helmholtz Centre for Environmental Research GmbH - UFZ, Department of Environmental Informatics, Leipzig, Germany

## Abstract

This paper discusses tension-compression asymmetry properties of Ogden hyperelastic formulations. It is shown that if all negative or all positive Ogden coefficients are used, tension-compression asymmetry occurs the degree of which cannot be separately controlled from the degree of non-linearity. A simple hybrid form is therefore proposed providing separate control over the tension-compression asymmetry. It is demonstrated how this form relates to a newly introduced generalised strain tensor class which encompasses both the tension-compression asymmetric Seth-Hill strain class and the tension-compression symmetric Bažant strain class. If the control parameter is set to $q = 0.5$ a tension-compression symmetric form involving Bažant strains is obtained with the property $\Psi(\lambda_1, \lambda_2, \lambda_3) = \Psi\left(\frac{1}{\lambda_1}, \frac{1}{\lambda_2}, \frac{1}{\lambda_3}\right)$. The symmetric form may be desirable for the definition of ground matrix contributions in soft tissue modelling allowing all deviation from the symmetry to stem solely from fibrous reinforcement. Such an application is also presented demonstrating the use of the proposed formulation in the modelling of the non-linear elastic and transversely isotropic behaviour of skeletal muscle tissue in compression (the model implementation and fitting procedure have been made freely available). The presented hyperelastic formulations may aid researchers in independently controlling the degree of tension-compression asymmetry from the degree of non-linearity, and in the case of anisotropic materials may assist in determining the role played by, either the ground matrix, or the fibrous reinforcing structures, in generating asymmetry.

# 1. Introduction

Realistic constitutive modelling for biological soft tissue is relevant to research areas such as impact biomechanics[1–3], rehabilitation engineering[4–7], tissue engineering[8], gait analysis[9], surgical simulation[10–12], and modelling of soft tissue drug transport[13,14]. Many biological materials present with significantly different behaviour for tensile or compressive loading (e.g. muscle tissue[15], cervical tissue[16], bone[17], intervertebral disk[18], and cartilage[19]). This is known as tension-compression asymmetry. Further, biological materials are also non-linear elastic, and often anisotropic due to the presence of fibrous connective tissue structures. Anisotropy can be modelled by combining an isotropic ground matrix with fibrous reinforcement. Through adjustment of material parameters current hyperelastic constitutive formulations offer control over the dominance of either the ground-matrix or the fibrous components, as well as the degree of non-linearity in their response. Tension-compression asymmetry can be present in the behaviour of isotropic formulations, and therefore ground-matrix formulations, as well as in fibrous reinforcement components. However, in many formulations the constitutive parameters dictating the degree of non-linearity also affect the degree of tension-compression asymmetry. As such the source and degree of tension-compression asymmetry is often not specifically controlled in the constitutive formulation. This study presents constitutive formulations, based on Ogden hyperelasticity, for isotropic materials, such as ground-matrices, offering separate control over the degree of non-linearity and tension-compression asymmetry. Such formulations may aid in the identification of the role played by either the ground matrix, or the fibrous reinforcing structures, in generating tension-compression asymmetry.

According to the representation theorems in non-linear continuum mechanics (see also [20–22]) for isotropic materials the strain energy density function defining constitutive behaviour for finite elasticity can be formulated in terms of principal invariants $\{I_1, I_2, I_3\}$ or principal stretches $\{\lambda_1, \lambda_2, \lambda_3\}$ such that $\Psi(I_1, I_2, I_3) = \Psi(\lambda_1, \lambda_2, \lambda_3)$. Several successful formulations have been proposed for incompressible rubber-like materials (see also[23]). A general first order expression for incompressible materials in terms of principal invariants is given by the so called Mooney-Rivlin hyperelastic model (Mooney 1940[24] and Rivlin 1948[25,26]) often presented as:

$$\Psi(I_1(\mathbf{C}), I_2(\mathbf{C})) = C_1(I_1(\mathbf{C}) - 3) + C_2(I_2(\mathbf{C}) - 3) \qquad 1$$

For incompressible materials, $J = \det(\mathbf{F}) = 1$ and therefore $I_2(\mathbf{C}) = I_1(\mathbf{C}^{-1})$, allowing the Mooney-Rivlin form to be rewritten in terms of principal stretches as:

$$\Psi(\lambda_1, \lambda_2, \lambda_3) = C_1(\lambda_1^2 + \lambda_2^2 + \lambda_3^2 - 3) + C_2(\lambda_1^{-2} + \lambda_2^{-2} + \lambda_3^{-2} - 3) \qquad 2$$

Mooney 1940[24] derived this form by postulating isotropy, incompressibility, and the requirement that tractions for simple shear are proportional to the shear. To capture non-linear behaviour for finite deformations Mooney 1940[24] proposed a more general form whereby tractions were postulated to be explicit functions of the shear:

$$\Psi(\lambda_1, \lambda_2, \lambda_3) = \sum_{m=1}^{\infty} \left[ A_{2m}(\lambda_1^{2m} + \lambda_2^{2m} + \lambda_3^{2m} - 3) + B_{2m}(\lambda_1^{-2m} + \lambda_2^{-2m} + \lambda_3^{-2m} - 3) \right] \qquad 3$$

(follows original notation by Mooney 1940[24], note that $m \in \mathbb{N}$, and plays the role of a subscript index, for the constitutive parameters, and appears in the exponent for the stretches). It can be seen that if $A_{2m} = B_{2m}$ this form has the tension and compression symmetry property $\Psi(\lambda_1, \lambda_2, \lambda_3) = \Psi\left(\frac{1}{\lambda_1}, \frac{1}{\lambda_2}, \frac{1}{\lambda_3}\right)$. Mooney 1940[24] and more specifically Rivlin 1948[25,26] argued the strain energy density should be a symmetrical and even-powered function of the principal stretches. Mooney 1940[24] also presented a form offering control over asymmetry by using $A_{2m} = \frac{G_{2m} + H_{2m}}{4}$ and $B_{2m} = \frac{G_{2m} - H_{2m}}{4}$:

$$\Psi(\lambda_1, \lambda_2, \lambda_3) = \sum_{m=1}^{\infty} \left[ \sum_{i=1}^{3} \frac{G_{2m}}{4m} \left(\lambda_i^{2m} + \lambda_i^{-2m} - 2\right) + \sum_{i=1}^{3} \frac{H_{2m}}{4m} \left(\lambda_i^{2m} - \lambda_i^{-2m}\right) \right] \qquad 4$$

Ogden 1972[27,28] also removed the symmetry constraint, and, since stretches are naturally positive quantities, dropped the requirement for integer and even-powers leading to the highly flexible form:

$$\Psi(\lambda_1, \lambda_2, \lambda_3) = \sum_{a=1}^{N} \frac{c_a}{m_a} \left(\lambda_1^{m_a} + \lambda_2^{m_a} + \lambda_3^{m_a} - 3\right) = \sum_{a=1}^{N} \left[ \frac{c_a}{m_a} \sum_{i=1}^{3} \left(\lambda_i^{m_a} - 1\right) \right] \qquad 5$$

With $m_a \in \mathbb{R}$, and $(c_a m_a) \in \mathbb{R}_{>0}$

The Ogden hyperelastic formulation has been employed to a great extent for incompressible rubber-like materials[29,30] and has recently been shown to agree with the statistical theory of microscopic fibre networks[31]. Typically for rubber-like materials parameter fitting of the Ogden form involves 3-4 terms, whereby 1 term involves negative $m_a$ and 2-3 terms involve positive $m_a$ values (see also[30–32]). Since mechanical testing of biological samples is more challenging, and the data is often of a sparser nature compared to data for engineering materials, reduced order models are often employed leaving fewer parameters to be identified. For instance, 1st order Ogden formulations have been used for skeletal muscle tissue[33] and skin[34]. In this case only positive $m_a$ values are used. For such reduced order formulations, as will be demonstrated in this paper, a tension-compression asymmetry exists. When the parameters controlling the degree of non-linearity (the $m_a$ values) are adjusted the asymmetry is also affected. Hence control of the degree of non-linearity, and the degree of asymmetry in tension and compression is not independent.

This study presents Ogden formulations in light of generalised strains including the Seth-Hill (Seth 1961[35] and Hill 1968[36]) and Bažant (Bažant 1998[37]) strain measures. The implications of the use of only positive or only negative $m_a$ values are discussed. It is demonstrated how the properties and shortcomings of the Seth-Hill strains are reflected in the model outcome. In particular it is shown how models with all positive or all negative $m$ values (which includes models for $N = 1$ and therefore the Neo-Hookean forms) lead to behaviour that is either fully defined by tension or compression processes, respectively. For instance, in the case of uniaxial loading, this means that the resistance to the load may not stem from direct resistance in the loading direction, but instead stems predominantly from the hydrostatic pressure and induced deformation processes orthogonal to the load direction (e.g. for uniaxial compression, and all positive $m$ values, resistance is dominated by induced tensile processes orthogonal to the compression direction, and the hydrostatic pressure).

A variation on the Ogden formulation is therefore proposed allowing for the control of tension-compression asymmetry and is shown to be related to a novel set of generalised strain tensors. A special case of the proposed formulation allows for the retrieval of the symmetry property $\Psi(\lambda_1, \lambda_2, \lambda_3) = \Psi\left(\frac{1}{\lambda_1}, \frac{1}{\lambda_2}, \frac{1}{\lambda_3}\right)$. It is shown how this formulation relates to the symmetric Bažant generalised strain measures which offer more desirable limits for finite strains than the Seth-Hill class. Such a symmetric formulation may be useful for the description of ground matrix contributions in fibre reinforced materials. This allows one to attribute a material's tension-compression asymmetry solely to its fibrous reinforcement. The more general form presented offers control over the asymmetry allowing for the more independent investigation of asymmetry and non-linear elasticity. In the case of anisotropic formulations this form enables the investigation of the role played by, either the ground matrix, or the fibrous reinforcing structures, in generating tension-compression asymmetry.

As an application of the proposed formulation, the non-linear elastic and transversely isotropic behaviour of skeletal muscle tissue in compression was modelled.

## 2. Generalised strain tensors

This section highlights deformation metrics treated in the paper and reviews generalised strain tensor sets.

The right-Cauchy-Green tensor is given by:

$$\mathbf{C} = \mathbf{U}^2 = \mathbf{F}^{\mathrm{T}}\mathbf{F} \qquad\qquad 6$$

With $\mathbf{F}$ the deformation gradient tensor and $\mathbf{U}$ the right stretch tensor. The eigenvalues or principal components of $\mathbf{C}$ are $C_i = \lambda_i^2$ (with $i = 1, 2, 3$), i.e. the squared principal stretches.

A general class of finite (Lagrangian) strain tensors is given by:

$$\mathbf{E}^{(m)} = \begin{cases} m \neq 0 & \frac{1}{m}(\mathbf{U}^m - \mathbf{I}) \\ m = 0 & \ln(\mathbf{U}) \end{cases} \qquad\qquad 7$$

This class of strain tensors is sometimes referred to as the Seth-Hill (Seth 1961[35] and Hill 1968[36]), or the Doyle-Ericksen[38] class. For instance for $m = 0$, $m = 1$, and $m = 2$ one obtains the Hencky, Biot and Green-Lagrange strain tensors respectively. If $m \neq 0$ the principal components of $\mathbf{E}^{(m)}$ are:

$$E_i^{(m)} = \frac{1}{m}(\lambda_i^m - 1) \qquad\qquad 8$$

, and the first invariant of $\mathbf{E}^{(m)}$ has the form:

$$I_1(\mathbf{E}^{(m)}) = \mathrm{tr}(\mathbf{E}^{(m)}) = \frac{1}{m}(\lambda_1^m + \lambda_2^m + \lambda_3^m - 3) = \frac{1}{m}\sum_{i=1}^{3}(\lambda_i^m - 1) \qquad\qquad 9$$

Although widely used, these strain measures exhibit some finite limits for extreme tension and compression which may not be desirable analytically. For instance for $m > 0$ we have the properties:

$$\lim_{\lambda_i \to \infty} E_i^{(m)} = \infty, \text{ and } \lim_{\lambda_i \to 0} E_i^{(m)} = -\frac{1}{m} \qquad\qquad 10$$

While for $m < 0$:

$$\lim_{\lambda_i \to \infty} E_i^{(m)} = \frac{1}{m}, \text{ and } \lim_{\lambda_i \to 0} E_i^{(m)} = -\infty \qquad 11$$

The only exception is the Hencky strain obtained for $m = 0$. Several authors have therefore investigated strain measures which offer a response which is effectively an average or hybrid of the two subclasses for $m < 0$ and $m > 0$. For instance Bažant 1998[37], and later Billington 2003[39] present:

$$\Xi^{(m)} = \frac{1}{2}\left(\mathbf{E}^{(m)} + \mathbf{E}^{(-m)}\right) = \frac{1}{2m}(\mathbf{U}^m - \mathbf{U}^{-m}) \qquad 12$$

The principal components of $\Xi^{(m)}$ are:

$$\Xi_i^{(m)} = \frac{1}{2m}(\lambda_i^m - \lambda_i^{-m}) \qquad 13$$

Therefore, the first invariant of $\Xi^{(m)}$ has the form:

$$I_1\left(\Xi^{(m)}\right) = \mathrm{tr}\left(\Xi^{(m)}\right) = \frac{1}{2m}\sum_{i=1}^{3}(\lambda_i^m - \lambda_i^{-m}) \qquad 14$$

The class $\Xi^{(m)}$ offers the properties:

$$\lim_{\lambda_i \to \infty} \Xi_i^{(m)} = \infty, \text{ and } \lim_{\lambda_i \to 0} \Xi_i^{(m)} = -\infty \qquad 15$$

These properties are shared with the Hencky strain and indeed Bažant 1998[37] shows these measures allow approximation of the Hencky strain tensor. In addition, the Bažant strains have the following symmetry property for tension and compression:

$$\Xi_i^{(m)}(\lambda_i) = -\Xi_i^{(m)}\left(\frac{1}{\lambda_i}\right) \qquad 16$$

Figure 1 illustrates the properties of several common strain measures, Seth-Hill strains, as well as a Bažant strain measure $\Xi^{(2)}$ (similar to [40]).

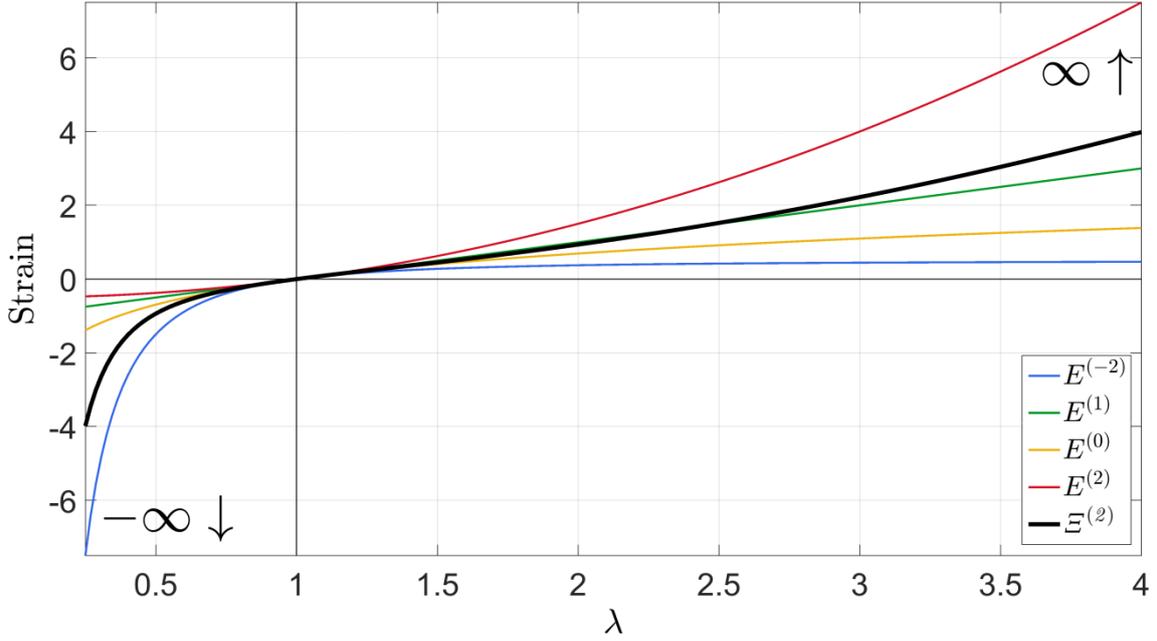

*Figure 1 Various strain measures as a function of stretch (uniaxial). The first 4 coloured curves represent Seth-Hill strains $E^{(m)}$, where the green, yellow and red represent the Biot, Henky and Green-Lagrange strains respectively. The black curve represents the Bažant strain $\Xi^{(2)}$ capable of spanning $\{-\infty, \infty\}$.*

For uncoupled formulations, deviatoric deformation measures are required. These can be derived using the volume ratio $J = \det(\mathbf{F})$:

$$\tilde{\mathbf{C}} = J^{-\frac{2}{3}}\mathbf{C}, \quad \tilde{\lambda}_i = J^{-\frac{1}{3}}\lambda_i, \quad \tilde{\mathbf{U}} = J^{-\frac{1}{3}}\mathbf{U} \qquad 17$$

Leading to the deviatoric generalised strains:

$$\tilde{\mathbf{E}}^{(m)} = \begin{cases} m \neq 0 & \frac{1}{m}(\tilde{\mathbf{U}}^m - \mathbf{I}) \\ m = 0 & \ln(\tilde{\mathbf{U}}) \end{cases} \qquad 18$$

$$\tilde{\Xi}^{(m)} = \frac{1}{2}\left(\tilde{\mathbf{E}}^{(m)} + \tilde{\mathbf{E}}^{(-m)}\right)$$

# 3. Ogden hyperelasticity in relation to invariants and generalised strains

Ogden 1972[27,28] (see also[22]) presented the strain energy density function presented in equation 5. The Cauchy stress $\boldsymbol{\sigma}$ for such a strain energy density function is given by[20]:

$$\boldsymbol{\sigma} = \bar{\boldsymbol{\sigma}} - \bar{p}\mathbf{I}, \quad \bar{\sigma}_i = \lambda_i \frac{\partial \Psi}{\partial \lambda_i} \qquad 19$$

The indeterminate variable $\bar{p}$ is part of the hydrostatic pressure and functions as a Lagrange multiplier of the incompressibility constraint. It is determined from the boundary conditions rather than a constitutive equation. The total hydrostatic pressure $p$ can be identified as:

$$p = -\frac{1}{3}\mathrm{tr}(\boldsymbol{\sigma}) \qquad 20$$

In some cases, the strain energy density is instead written as $\Psi = \Psi(\lambda_1, \lambda_2, \lambda_3) - \bar{p}(J-1)$ highlighting the role of $\bar{p}$, however for constrained formulations it is here implied by the constraints in the entropy inequality rather than defined in the constitutive form.

Some Ogden implementations use $\frac{c_a}{m_a^2}$ (e.g. [41]) instead of $\frac{c_a}{m_a}$ leading to:

$$\Psi(\lambda_1, \lambda_2, \lambda_3) = \sum_{a=1}^{N}\left[\frac{c_a}{m_a^2}\sum_{i=1}^{3}(\lambda_i^{m_a} - 1)\right] \quad 21$$

With $m_a \in \mathbb{R}$, and $c_a \in \mathbb{R}_{>0}$

The latter forms the focus of this paper and has been implemented in the open source finite element code FEBio[42] (v2.1.1, Musculoskeletal Research Laboratories, The University of Utah, USA). From equation 9 it is clear that one may recast the Ogden form in terms of traces, or first invariants, of generalised Seth-Hill strain tensors (see also [22,27,28]):

$$\Psi(\mathbf{E}^{(m_a)}) = \sum_{a=1}^{N}\frac{c_a}{m_a}\text{tr}(\mathbf{E}^{(m_a)}) = \sum_{a=1}^{N}\frac{c_a}{m_a}I_1(\mathbf{E}^{(m_a)}) \quad 22$$

The above is for the incompressible and constrained Ogden forms. An unconstrained or coupled variant of the form in equation 21 is given by (see also[22]):

$$\Psi = \frac{\kappa'}{2}(J-1)^2 + \sum_{a=1}^{N}\left[\frac{c_a}{m_a^2}\left(\left(\sum_{i=1}^{3}(\lambda_i^{m_a} - 1)\right) - m_a \ln(J)\right)\right]$$

$$= \frac{\kappa'}{2}(J-1)^2 + \sum_{a=1}^{N}\frac{c_a}{m_a}\left(\text{tr}(\mathbf{E}^{(m_a)}) - \ln(J)\right) \quad 23$$

With $m_a \in \mathbb{R}$, and $c_a, \kappa' \in \mathbb{R}_{>0}$

Here $\kappa'$ is a material parameter similar to a bulk-modulus.

When dealing with nearly incompressible materials, it is convenient, for numerical implementation, to uncouple the strain energy density function into its isochoric (deviatoric) and volumetric parts denoted $\Psi_{\text{iso}}(\tilde{\lambda}_1, \tilde{\lambda}_2, \tilde{\lambda}_3)$ and $\Psi_{\text{vol}}(J)$ respectively, leading to $\Psi = \Psi_{\text{iso}}(\tilde{\lambda}_1, \tilde{\lambda}_2, \tilde{\lambda}_3) + \Psi_{\text{vol}}(J)$. An uncoupled variant of equation 21 is:

$$\Psi_{\text{iso}} = \sum_{a=1}^{N}\left[\frac{c_a}{m_a^2}\sum_{i=1}^{3}(\tilde{\lambda}_i^{m_a} - 1)\right] = \sum_{a=1}^{N}\frac{c_a}{m_a}\text{tr}(\tilde{\mathbf{E}}^{(m_a)})$$

$$\Psi_{\text{vol}} = \frac{\kappa}{2}\ln(J)^2 \quad 24$$

With $m_a \in \mathbb{R}$, and $c_a, \kappa \in \mathbb{R}_{>0}$

(various forms of $\Psi_{\text{vol}}(J)$ have been proposed, the presented form is implemented in FEBio[42]). In the above, the material parameter $\kappa$ represents the bulk modulus.

# 4. Tension-Compression behaviour in Ogden hyperelasticity

From the preceding section it has become clear that the constrained, unconstrained and uncoupled Ogden formulations can all be expressed as functions of Seth-Hill generalised strains (and the volume ratio $J$). This section will focus on tension-compression symmetry or asymmetry in Ogden hyperelasticity. The constrained formulation will be used to guide the discussion; however the arguments can be extended to the unconstrained and uncoupled formulations as well.

The principal Cauchy stresses for the constrained and incompressible Ogden form (equation 21) become:

$$\sigma_i = -\bar{p} + \sum_{a=1}^{N} \frac{c_a}{m_a} \lambda_i^{m_a} = -\bar{p} + \sum_{a=1}^{N} \left( c_a E_i^{(m_a)} + \frac{c_a}{m_a} \right) \qquad 25$$

In the reference state $\lambda_j = 1$ and we require $\sigma_i = 0$ leading to:

$$\sigma_i(\lambda_j = 1) = -\bar{p} + \bar{\sigma}_i = 0$$

$$\bar{\sigma}_i(\lambda_j = 1) = \sum_{a=1}^{N} \frac{c_a}{m_a} \;\to\; \bar{p} = -\bar{\sigma}_i(\lambda_j = 1) \;\to\; p = 0 \qquad 26$$

If only negative or only positive $m$ values are employed, $\bar{\sigma}_i$ and $\bar{p}$ are therefore non-zero in the reference configuration.

As discussed in section 2, negative or positive $m_a$ values cause the Seth-Hill measures to be most sensitive to compression or tension respectively. Since the Ogden formulations involve Seth-Hill measures this behaviour is also reflected in the constitutive law. Figure 2 illustrates the behaviour for uniaxial loading in the 1-direction with the conditions:

$$J = \lambda_1 \lambda_2 \lambda_3 = 1, \quad \lambda_2 = \lambda_3 = \lambda_1^{-\frac{1}{2}}, \quad \sigma_2 = \sigma_3 = 0$$

$$\to \bar{p} = \sum_{a=1}^{N} \frac{c_a}{m_a} \lambda_2^{m_a} = \sum_{a=1}^{N} \frac{c_a}{m_a} \left( \lambda_1^{-\frac{1}{2}} \right)^{m_a} = \sum_{a=1}^{N} \frac{c_a}{m_a} \lambda_1^{-\frac{m_a}{2}} \qquad 27$$

$$\to \sigma_1 = \sum_{a=1}^{N} \frac{c_a}{m_a} \left( \lambda_1^{m_a} - \lambda_1^{-\frac{m_a}{2}} \right)$$

The solid blue and red curves in Figure 2 show the typical behaviour when only negative or only positive $m_a$ values are employed, respectively. It can be seen that the former is most sensitive to compression while the latter is most sensitive to tension. The curves are plotted for $N = 1$, however the asymmetry property is maintained for $N > 1$ if only negative or only positive $m_a$ values are employed. The other curves are for the proposed hybrid formulation which will be discussed in chapter 5.

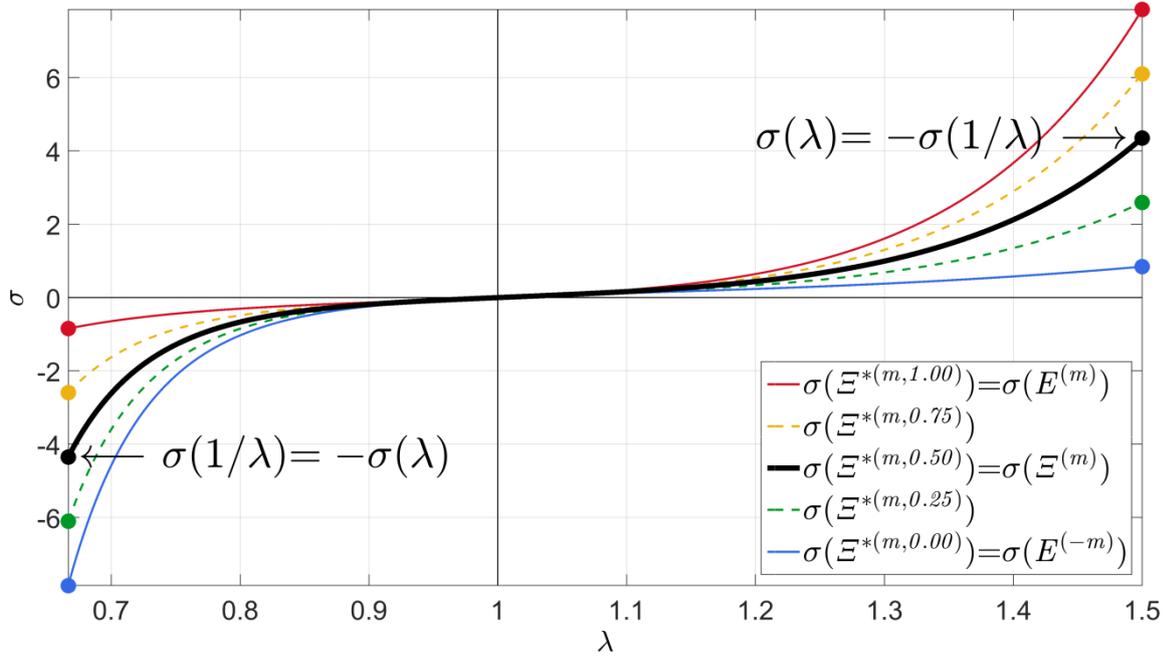

*Figure 2 Typical uniaxial stress responses for Ogden formulations with $m_a < 0$ (blue), $m_a > 0$ (red). Hybrid forms are also shown in yellow and green and, in the middle, a form (black) with the tension-compression symmetry property $\sigma(\lambda) = -\sigma(1/\lambda)$. The legend highlights the relationship between the stresses and the strain measures. All curves were created for N=1, $m = 11$, and $c = 1$. The strain tensors $\Xi^{*(m,q)}$ are clarified in chapter 5 equation 36.*

First of all it is clear that for all formulations the following basic requirements are fulfilled:

$$\sigma_i(\lambda_j = 1) = 0, \quad \lim_{\lambda_i \to \infty} \sigma_i(\lambda_i) = \infty, \quad \lim_{\lambda_i \to 0} \sigma_i(\lambda_i) = -\infty \qquad 28$$

However, a delayed (with respect to stretch) onset in the tension or compression regime is observed for formulations employing only negative or only positive $m_a$ values respectively. This is due to the nature of the "resistive processes" implied by either type of choice of $m_a$ value. These resistive processes in relation to $m_a$ values are schematically illustrated in Figure 3 as springs, and will now be discussed in more detail.

If all $m_a > 0$ we see:

$$\bar{\sigma}_i(\lambda_i) = \begin{cases} > 0 & \lambda_i > 1 \\ \geq 0 & \lambda_i < 1 \end{cases}, \quad \lim_{\lambda_i \to \infty} \bar{\sigma}_i(\lambda_i) = \infty, \quad \lim_{\lambda_i \to 0} \bar{\sigma}_i(\lambda_i) = 0 \qquad 29$$

If all $m_a > 0$ then for compression with $\lambda_1 < 1$ the stress contribution $\bar{\sigma}_1$ is either zero or of a tensile nature. For the example of uniaxial loading (e.g. red curve in Figure 2) a negative decreasing stress does appropriately develop in compression due to $\lambda_1$, but only due to the dominance of the effect of the Lagrangian multiplier $\bar{p}$ which grows monotonically due to the tensile processes in $\lambda_2 = \lambda_3 = \lambda_1^{-\frac{1}{2}} > 1$. Therefore, if all $m_a > 0$, the entire model behaviour, whether in tension or compression, is dictated by direct or induced tensile processes. In Figure 3 these tensile contributions are highlighted as red springs.

The situation is reversed for the case of all $m_a < 0$ where we observe:

$$\bar{\sigma}_i(\lambda_i) = \begin{cases} \leq 0 & \lambda_i > 1 \\ < 0 & \lambda_i < 1 \end{cases}, \quad \lim_{\lambda_i \to \infty} \bar{\sigma}_i(\lambda_i) = 0, \quad \lim_{\lambda_i \to 0} \bar{\sigma}_i(\lambda_i) = -\infty \qquad 30$$

Now if $m_a < 0$ then for tension with $\lambda_1 > 1$ the stress contribution $\bar{\sigma}_1$ is either zero or of a compressive nature. For the example of uniaxial loading (e.g. blue curve in Figure 2) a positive increasing stress does appropriately develop in tension due to $\lambda_1$ but again only due to the dominance of the effect of the Lagrangian multiplier $\bar{p}$ which decreases monotonically due to the compressive processes in $\lambda_2 = \lambda_3 = \lambda_1^{-\frac{1}{2}} < 1$. Therefore, if all $m_a < 0$, the entire model behaviour, whether in tension or compression, is dictated by direct or induced compressive processes. In Figure 3 these compressive contributions are highlighted as blue springs. For incompressible behaviour and/or isochoric conditions whereby $J = 1$, it is noted that $I_1(\mathbf{C}^{-m}) = I_2(\mathbf{C}^m)$. In that case what is here termed "compressive resistive processes" is sometimes referred to in the literature as a resistance relatable to "tube-like constraints" (e.g. [31]) or area changes.

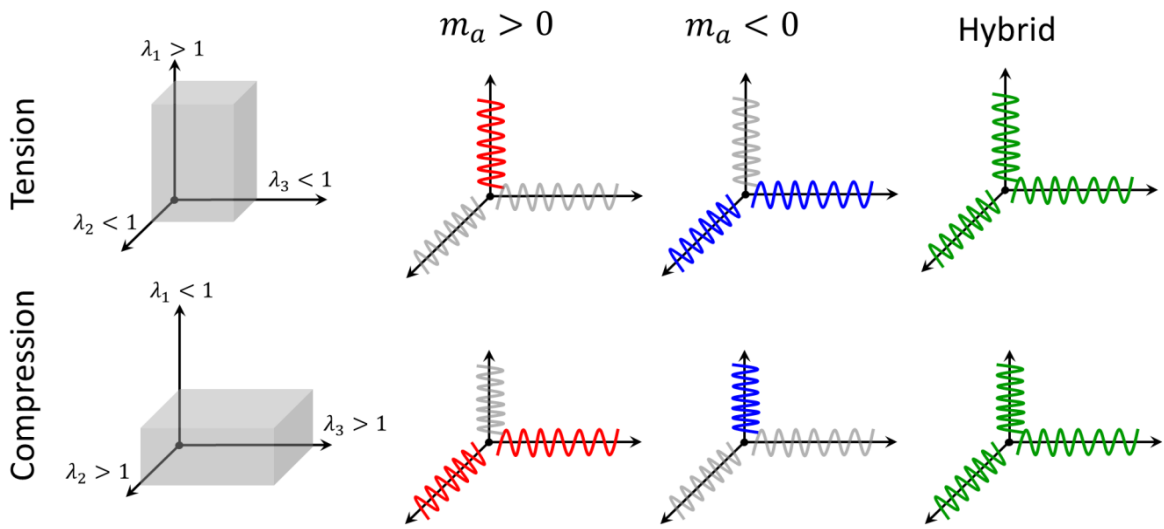

*Figure 3 Schematic illustration of resistive processes during uniaxial loading with the top and bottom row representing tension and compression respectively. Red and blue springs denote contributions for $m_a > 0$ and $m_a < 0$ respectively. Green springs denote contributions for a hybrid formulation with both positive and negative $m_a$. Gray springs denote diminishing contributions and may be of a tensile or compressive nature despite stretches being of a compressive or tensile nature respectively.*

## 5. The proposed formulation with control of asymmetry

The previous section demonstrated that formulations with equal signs for all $m_a$ exhibit a tension-compression asymmetry since the increase of stresses is stalled in one of the loading domains with respect to the other. These formulations are by no means invalid. The asymmetric nature may be demonstrable from experiments for certain materials. However, for these formulations the degree of non-linearity and the degree of asymmetry are both affected by the $m_a$ parameters. Hence independent investigation of these phenomena may be challenging. This chapter therefore presents an Ogden formulation offering more independent control of non-linearity and tension-compression asymmetry. In addition, a special case of this form is discussed enforcing tension-compression symmetry.

An alternative Ogden formulation (with an incompressibility constraint) can be composed whereby the sum of second order forms is enforced i.e.:

$$\Psi(\lambda_1, \lambda_2, \lambda_3) = \sum_{a=1}^{N} \left[ \frac{c_{1a}}{m_{1a}^2} \left( \sum_{i=1}^{3} (\lambda_i^{m_{1a}} - 1) \right) + \frac{c_{2a}}{m_{2a}^2} \left( \sum_{j=1}^{3} (\lambda_j^{m_{2a}} - 1) \right) \right] \quad 31$$

With $c_{1a}, c_{2a}, m_{1a} \in \mathbb{R}_{>0}$ and $m_{1a} = -m_{2a}$

Where $m_{1a} = -m_{2a}$ is used in order to enforce the use of matched positive and negative $m$ parameters. Employing this constraint results in a form similar to the Mooney formulation described by equation 3. In fact if $N = 1$ and $m_{11} = -m_{21} = 2$ the following Mooney-Rivlin form is obtained:

$$\Psi(\lambda_1, \lambda_2, \lambda_3) = \frac{c_{11}}{4}(\lambda_1^2 + \lambda_2^2 + \lambda_3^2 - 3) + \frac{c_{21}}{4}(\lambda_1^{-2} + \lambda_2^{-2} + \lambda_3^{-2} - 3) \quad 32$$

If the tension-compression symmetry property $\Psi(\lambda_1, \lambda_2, \lambda_3) = \Psi\left(\frac{1}{\lambda_1}, \frac{1}{\lambda_2}, \frac{1}{\lambda_3}\right)$ is desired the additional constraint $c_{1a} = c_{2a} = c_a$ is sufficient. If instead some control over the tension-compression asymmetry is required one may return to equation 31, and analogous to equation 4, employ $c_{1a} = \frac{G_a + H_a}{4}$ and $c_{2a} = \frac{G_a - H_a}{4}$. An alternative means of controlling asymmetry, which is favoured here, is obtained by introducing a weighting factor $q$ and using $c_{1a} = qc_a$, $c_{2a} = (1-q)c_a$, $m_{1a} = -m_{2a} = m_a$, leading to:

$$\Psi(\lambda_1, \lambda_2, \lambda_3) = \sum_{a=1}^{N} \frac{c_a}{m_a} \left[ q \frac{1}{m_a} \left( \sum_{i=1}^{3} (\lambda_i^{m_a} - 1) \right) - (1-q) \frac{1}{-m_a} \left( \sum_{j=1}^{3} (\lambda_j^{-m_a} - 1) \right) \right] \quad 33$$

With $q \in [0,1]$, $c_a, m_a \in \mathbb{R}_{>0}$

Note that the minus signs were deliberately not cancelled out such that it is clear that through the definition of $\mathbf{E}^{(-m_a)}$, and making use of equation 9, this can be rewritten:

$$\Psi = \sum_{a=1}^{N} \frac{c_a}{m_a} \left( q \operatorname{tr}(\mathbf{E}^{(m_a)}) - (1-q) \operatorname{tr}(\mathbf{E}^{(-m_a)}) \right) \quad 34$$

With $q \in [0,1]$, $c_a, m_a \in \mathbb{R}_{>0}$

This is the most general form of the proposed formulation. The principal Cauchy stress for this formulation can be written:

$$\bar{\sigma}_i = \sum_{a=1}^{N} c_a \left( qE_i^{(m_a)} + (1-q)E_i^{(-m_a)} + \frac{2q-1}{m_a} \right) = \sum_{a=1}^{N} c_a \left( \Xi_i^{*(m_a, q)} + \frac{2q-1}{m_a} \right) \quad 35$$

Here, $\Xi^{*(m_a, q)}$ is used to introduce the novel generalised strain set:

$$\Xi^{*(m,q)} = q\mathbf{E}^{(m)} + (1-q)\mathbf{E}^{(-m)}$$

$$\Xi_i^{*(m,q)} = q\left(\frac{1}{m}(\lambda_i^m - 1)\right) + (1-q)\left(\frac{1}{m}(1 - \lambda_i^{-m})\right) \quad 36$$

$$q \in [0,1]$$

Observing Figure 4 and the particular examples:

$$q = 1 \rightarrow \Xi^{*(m,1)} = \mathbf{E}^{(m)}$$
$$q = \frac{1}{2} \rightarrow \Xi^{*(m,\frac{1}{2})} = \Xi^{(m)} \qquad 37$$
$$q = 0 \rightarrow \Xi^{*(m,0)} = \mathbf{E}^{(-m)}$$

it becomes clear that the class $\Xi^{*(m,q)}$ contains both the Seth-Hill (i.e. if $q = 1$ or $q = 0$) and Bažant (i.e. if $q = 0.5$) classes as well as hybrid intermediate forms (if $0 < q < 1$). The limits for finite deformations for the Seth-Hill class (if $q = 0$ or $q = 1$) are shown in equations 10 and 11. For $0 < q < 1$ the limits are shared with the Bažant class which appear in 15.

The above demonstrates that $q$ is a weighting factor controlling sensitivity to the tension biased Seth-Hill strains $\mathbf{E}^{(m_a)}$ or the compression biased Seth-Hill strains $\mathbf{E}^{(-m_a)}$. In the special case $q = 0.5$ the two balance each other causing equation 34 to have the tension-compression property $\Psi(\lambda_1, \lambda_2, \lambda_3) = \Psi\left(\frac{1}{\lambda_1}, \frac{1}{\lambda_2}, \frac{1}{\lambda_3}\right)$. This is also reflected in the principal Cauchy stress expressions (see also Figure 2), e.g., for $q = 1$, $q = 0.5$, and $q = 0$ one finds:

$$q = 1 \rightarrow \quad \bar{\sigma}_i = \sum_{a=1}^{N} c_a \left( E_i^{(m_a)} + \frac{1}{m_a} \right)$$
$$q = \frac{1}{2} \rightarrow \quad \bar{\sigma}_i = \sum_{a=1}^{N} c_a \Xi_i^{(m_a)} \qquad 38$$
$$q = 0 \rightarrow \quad \bar{\sigma}_i = \sum_{a=1}^{N} c_a \left( E_i^{(-m_a)} - \frac{1}{m_a} \right)$$

Where it can be seen that for $q = 0.5$ the principal Cauchy stress is a simple function of the tension-compression symmetric Bažant strains leading to: $\bar{\sigma}_i(\lambda_i) = -\bar{\sigma}_i\left(-\frac{1}{\lambda_i}\right)$.

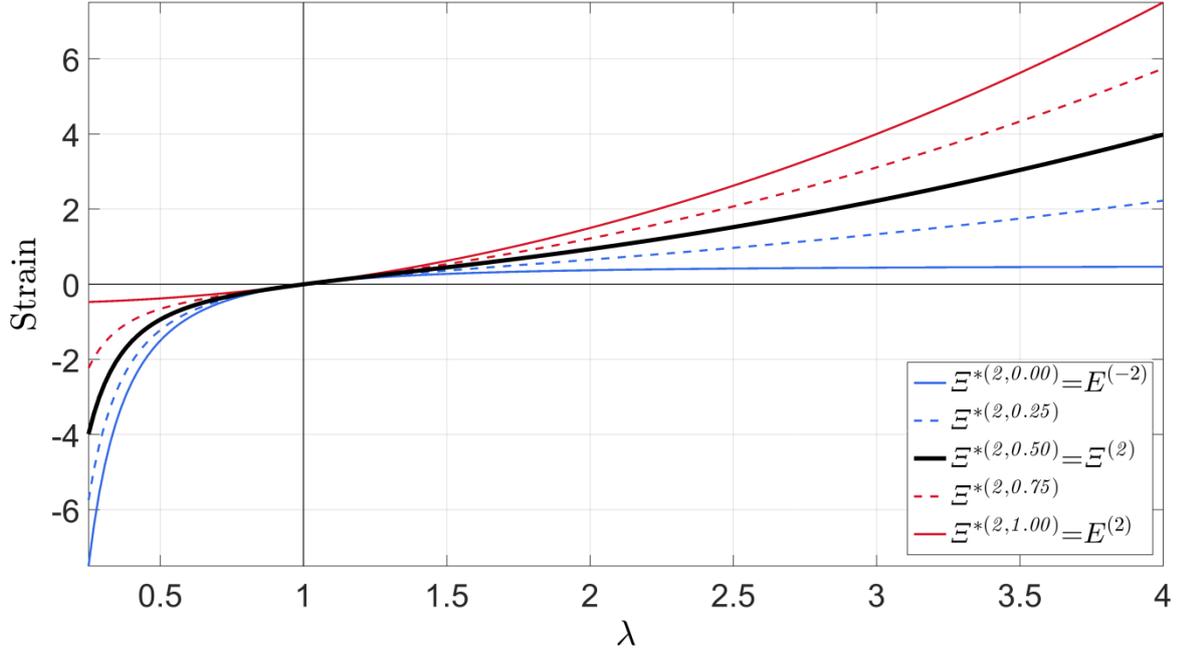

Figure 4 Uniaxial strain measures as a function of stretch for the proposed class $\Xi^{*(m,q)}$. Examples are shown for $m = 2$. Depending on the parameter $q$ the tension-compression asymmetric Seth-Hill strains or the symmetric Bazant strains can be retrieved. Dashed curves show intermediate and hybrid responses.

The properties of the form in equation 34 will now be discussed in more detail. The behaviour for $q = 0$ and $q = 1$ follows that discussed in chapter 4 for tension-compression asymmetric formulations with only negative or only positive $m$ values. The remainder of this chapter therefore focusses on the cases where $0 < q < 1$. In the reference configuration we have $\sigma_i(\lambda_j = 1) = 0$, and the following is observed:

$$\bar{\sigma}_i(\lambda_j = 1) = -\bar{p} = \sum_{a=1}^{N} c_a \frac{2q-1}{m_a} = \begin{cases} > 0 & 0.5 < q < 1 \\ 0 & q = 0.5 \\ < 0 & 0 < q < 0.5 \end{cases} \qquad 39$$

For non-unity stretches and $0 < q < 1$ we see:

$$\lim_{\lambda_i \to \infty} \bar{\sigma}_i(\lambda_i) = \infty, \quad \lim_{\lambda_i \to 0} \bar{\sigma}_i(\lambda_i) = -\infty \qquad 40$$

And the controllable tension-compression properties

$$\bar{\sigma}_i(\lambda_i) = \begin{cases} > -\bar{\sigma}_i\left(-\frac{1}{\lambda_i}\right) & 0.5 < q < 1 \\ -\bar{\sigma}_i\left(-\frac{1}{\lambda_i}\right) & q = 0.5 \\ < -\bar{\sigma}_i\left(-\frac{1}{\lambda_i}\right) & 0 < q < 0.5 \end{cases} \qquad 41$$

If $q = 0.5$ in the reference configuration, we see $\sigma_i(\lambda_j = 1) = \bar{\sigma}_i(\lambda_j = 1) = \bar{p} = p = 0$. For non-unity stretches we have symmetric behaviour and the desired properties:

$$\bar{\sigma}_i(\lambda_i) = -\bar{\sigma}_i\left(-\frac{1}{\lambda_i}\right), \quad \bar{\sigma}_i(\lambda_i) = \begin{cases} > 0 & \lambda_i > 1 \\ < 0 & \lambda_i < 1 \end{cases}$$
$$\lim_{\lambda_i \to \infty} \bar{\sigma}_i(\lambda_i) = \infty, \quad \lim_{\lambda_i \to 0} \bar{\sigma}_i(\lambda_i) = -\infty \qquad 42$$

Principal Cauchy stresses for the uniaxial loading example follow from:

$$\sigma_2 = \sigma_3 = -\bar{p} + \sum_{a=1}^{N} \frac{c_a}{m_a}\left(q\lambda_1^{-\frac{m_a}{2}} - (1-q)\lambda_1^{\frac{m_a}{2}}\right) = 0$$

$$\rightarrow \bar{p} = \sum_{a=1}^{N} \frac{c_a}{m_a}\left(q\lambda_1^{-\frac{m_a}{2}} - (1-q)\lambda_1^{\frac{m_a}{2}}\right) \qquad 43$$

$$\sigma_1 = \sum_{a=1}^{N} \frac{c_a}{m_a}\left[q\left(\lambda_1^{m_a} - \lambda_1^{-\frac{m_a}{2}}\right) - (1-q)\left(\lambda_1^{-m_a} - \lambda_1^{\frac{m_a}{2}}\right)\right]$$

Typical Cauchy stress behaviour for formulations with various $q$ values is shown in Figure 2. It can be seen that if $0 < q < 1$, a hybrid form is obtained featuring the sum of the behaviours due to positive and negative $m_a$ values. As a result, for these forms the response in tension or compression features direct tension and compression resistance. Such behaviour is illustrated by the green springs in Figure 2. If $q = 0.5$ is used, tension-compression symmetry is observed (black curve in Figure 2) owing to the properties of the strain measures $\Xi^{(m)}$.

For completeness, we now also present an unconstrained (coupled) and uncoupled variant of the form in equation 34. The former is given by:

$$\Psi = \frac{\kappa'}{2}(J-1)^2 + \sum_{a=1}^{N} \frac{c_a}{m_a}\left[q\text{tr}(\mathbf{E}^{(m_a)}) - (1-q)\text{tr}(\mathbf{E}^{(-m_a)}) + (1-2q)\ln(J)\right] \qquad 44$$

With $q \in [0,1]$, $c_a, m_a, \kappa' \in \mathbb{R}_{>0}$

And the principal Cauchy stresses can be evaluated from:

$$\sigma_i = J^{-1}\lambda_i \frac{\partial \Psi}{\partial \lambda_i}$$

$$\sigma_i = \kappa'(J-1) + J^{-1}\sum_{a=1}^{N} c_a\left(qE_i^{(m_a)} + (1-q)E_i^{(-m_a)}\right) = \kappa'(J-1) + J^{-1}\sum_{a=1}^{N} c_a\Xi_i^{*(m_a,q)} \qquad 45$$

An uncoupled variant of the form in equation 34 may employ the deviatoric variant of $\Xi^{*(m,q)}$ namely:

$$\widetilde{\Xi}^{*(m,q)} = q\widetilde{\mathbf{E}}^{(m)} + (1-q)\widetilde{\mathbf{E}}^{(-m)} \qquad 46$$

Leading to:

$$\Psi_{\text{iso}} = \sum_{a=1}^{N} \frac{c_a}{m_a}\left(q\text{tr}(\widetilde{\mathbf{E}}^{(m_a)}) - (1-q)\text{tr}(\widetilde{\mathbf{E}}^{(-m_a)})\right)$$

$$\Psi_{\text{vol}} = \frac{\kappa}{2}\ln(J)^2 \qquad 47$$

With $c_a, m_a, \kappa \in \mathbb{R}_{>0}$

From which the principal Cauchy stresses, with $\boldsymbol{\sigma} = \boldsymbol{\sigma}_{\text{iso}} + \boldsymbol{\sigma}_{\text{vol}}$, can be evaluated as:

$$\sigma_{\text{iso}_i} = J^{-1}\lambda_i \frac{\partial \Psi_{\text{iso}}}{\partial \lambda_i} = J^{-1}\left(\tilde{\lambda}_i \frac{\partial \Psi_{\text{iso}}}{\partial \tilde{\lambda}_i} - \frac{1}{3}\sum_{j=1}^{3} \tilde{\lambda}_j \frac{\partial \Psi_{\text{iso}}}{\partial \tilde{\lambda}_j}\right)$$

$$\sigma_{\text{iso}_i} = J^{-1}\sum_{a=1}^{N} c_a \left[\widetilde{\Xi}_i^{*(m_a,q)} - \frac{1}{3}\sum_{j=1}^{3}\left(\widetilde{\Xi}_j^{*(m_a,q)}\right)\right] \qquad 48$$

$$\sigma_{\text{vol}_i} = \kappa \frac{\ln(J)}{J}$$

# 6. Application to modelling skeletal muscle in compression

An application of the proposed formulation is now presented for modelling of the behaviour of skeletal muscle tissue in compression based on the data by Van Loocke et al. 2006[43]. The data represents quasi-static compression experiments (up to 30%) on passive, and freshly excised porcine skeletal muscle tissue. Figure 5A shows a schematic of a tissue sample with fibres and clarifies the Poisson's ratios found in that study, i.e.: $v_{31} = v_{32} = 0.5$, $v_{12} = v_{21} = 0.65$, $v_{13} = v_{23} = 0.36$. For fibre loading it was observed that in the final state the lateral stretches are simply $\lambda_2 = \lambda_3 = \lambda_1^{-\frac{1}{2}}$, while for cross-fibre loading the induced lateral stretches are $\lambda_2 = e^{-v_{12} log(\lambda_1 = 0.7)} \approx 1.261$ and $\lambda_3 = e^{-v_{13} log(\lambda_1 = 0.7)} \approx 1.137$. Figure 5B shows stress-stretch data for loading in the fibre and cross-fibre directions. The stresses are higher in the latter case perhaps due to the fact that reinforcing structures aligned with the muscle fibre direction may buckle for fibre direction loading. Van Loocke presented the use of an expansion of Hooke's law for a transversely isotropic material with strain dependant Young's Moduli to attempt to model the behaviour. However, the model does not respect the constraints for transverse isotropy ($\frac{v_{ij}}{E_i} = \frac{v_{ji}}{E_j}$, with $i \neq j$) and evaluation was based on closed form equations and fully prescribed experimental displacements. As such the evaluation presented here represents the first model fitting for the data by Van Loocke et al. 2006[43] using a valid transversely isotropic model.

In order to capture the non-linear elastic and transversely isotropic behaviour of skeletal muscle tissue the following general strain energy density form is used:

$$\Psi = \Psi_G + \Psi_F \qquad \qquad 49$$

Here $\Psi_G$ and $\Psi_F$ describe the contributions by the ground matrix and fibrous reinforcement respectively. The ground matrix is here defined by a first order variant of the proposed symmetric and unconstrained (coupled) formulation (equation 44 with $N = 1$):

$$\Psi_G = \frac{c}{2m} \left( \text{tr}(\mathbf{E}^{(m)}) - \text{tr}(\mathbf{E}^{(-m)}) \right) + \frac{\kappa'}{2}(J - 1)^2 \qquad \qquad 50$$

With $c, m, \kappa' \in \mathbb{R}_{>0}$

An unconstrained and coupled formulation is used here to ensure realistic stresses and deformations even for applications where $J \neq 1$ (see [44–46]).

Fibrous reinforcement was modelled by an ellipsoidal fibre distribution (for a more detailed discussion see [47]). A spherical distribution of fibres $\mathbf{n}_i$ is defined in a local orthonormal basis system $\mathcal{A} = \{\mathbf{a}_1, \mathbf{a}_2, \mathbf{a}_3\}$, with in this case $\mathbf{a}_3$ the fibre direction, and $\mathbf{a}_2 = \mathbf{e}_2$. In an associated spherical coordinate system, with angles $(\Theta, \Phi)$, the fibre vectors $\mathbf{n}_i$ can be represented as:

$$\mathbf{n}_i = \cos(\Theta_i) \sin(\Phi_i) \mathbf{a}_1 + \sin(\Theta_i) \sin(\Phi_i) \mathbf{a}_2 + \cos(\Phi_i) \mathbf{a}_3 \qquad \qquad 51$$

The strain energy density $\Psi_{F_i}$ for each of the fibres is defined as:

$$\Psi_{F_i}(\mathbf{n}_i, \mathbf{C}) = H(\lambda_{F_i} - 1)\xi(\mathbf{n}_i)(\lambda_{F_i}^2 - 1)^\beta, \quad \beta \geq 2 \qquad 52$$

Here, $\lambda_{F_i} = \sqrt{\mathbf{n}_i \cdot \mathbf{C}\mathbf{n}_i}$ is the stretch along the fibre direction $\mathbf{n}_i$, and $H$ is the Heaviside step function ensuring tension only contributions. The material parameter $\beta$ controls the degree of non-linearity and the parameters $\xi(\mathbf{n}_i)$ vary with fibre orientation according to the ellipsoidal function:

$$\xi(\mathbf{n}_i) = \left(\frac{\cos(\Theta_i)^2 \sin(\Phi_i)^2}{\xi_T^2} + \frac{\sin(\Theta_i)^2 \sin(\Phi_i)^2}{\xi_T^2} + \frac{\cos(\Phi_i)^2}{\xi_L^2}\right)^{-\frac{1}{2}} \qquad 53$$

Here $\xi_T$ and $\xi_L$ are fibre material parameters (units of stress) for the transverse and longitudinal direction, respectively.

The ground matrix response has the tension-compression symmetry property $\Psi_G(\lambda_1, \lambda_2, \lambda_3) = \Psi_G\left(\frac{1}{\lambda_1}, \frac{1}{\lambda_2}, \frac{1}{\lambda_3}\right)$, deviation from this symmetry for $\Psi$ occurs solely due to $\Psi_F$ which contains fibrous reinforcement defined for extensional processes.

The material parameters $\{c, m, \beta, \xi_T, \xi_L\}$ were determined using inverse finite element analysis (FEA) based on the open-source MATLAB toolbox GIBBON (r89, [48,49], http://www.gibboncode.org/). The bulk-modulus like parameter $\kappa'$ was constrained to be several orders of magnitude higher than the (more deviatoric) material parameters, i.e. $500(2c + \xi_T + \xi_L)$, resulting in $J = 1$. FEA was performed using the open source finite element software FEBio[42] (V2.2.6, Musculoskeletal Research Laboratories, The University of Utah, USA, http://febio.org/). A cubic sample (10x10x10 mm) was modelled using 125 hexahedral elements. One of the side faces, the front face, and the bottom face were constrained to not displace orthogonal to their surface. Compression was modelled by prescribing the motion orthogonal to the top face. Inverse parameter identification aimed to minimize the difference between experimental and numerically obtained stresses and deformations (mean absolute relative stress difference, and the mean absolute stretch differences), and employs Levenberg-Marquardt based optimisation (implemented using the MATLAB *lsqnonlin* function, see also [50]). The resulting model fits are overlain in Figure 5B. The goodness of fit is evident from $R^2$ values for the model response compared to the experimental data which were 0.995 and 0.999 for fibre and cross-fibre direction respectively. The fitted parameters were: $c = 0.6115$ kPa, $m = 2.007, \beta = 3.294, \xi_T = 0.09059$ kPa, $\xi_L = 21.30$ kPa. The mean and standard deviation of the error between the predicted and simulated stresses were -0.002 kPa and 0.012 kPa for fibre direction loading, and -0.006 kPa and 0.004 kPa for cross-fibre loading. Deviations in displacement were zero for fibre direction loading while for cross fibre loading the expansions in the cross-fibre and fibre directions could be matched to within 0.01 and 0.05 mm respectively. It is noted here that the deformations could not be matched (data not presented) using the more standard approach of reinforcing only along the fibre direction (instead of an ellipsoidal fibre distribution acting in all directions).

The model fitting procedures and implementation with FEBio have been made freely available in GIBBON and can be found in the example *DEMO_FEBio_iFEA_uniaxial_transiso_02.m*.

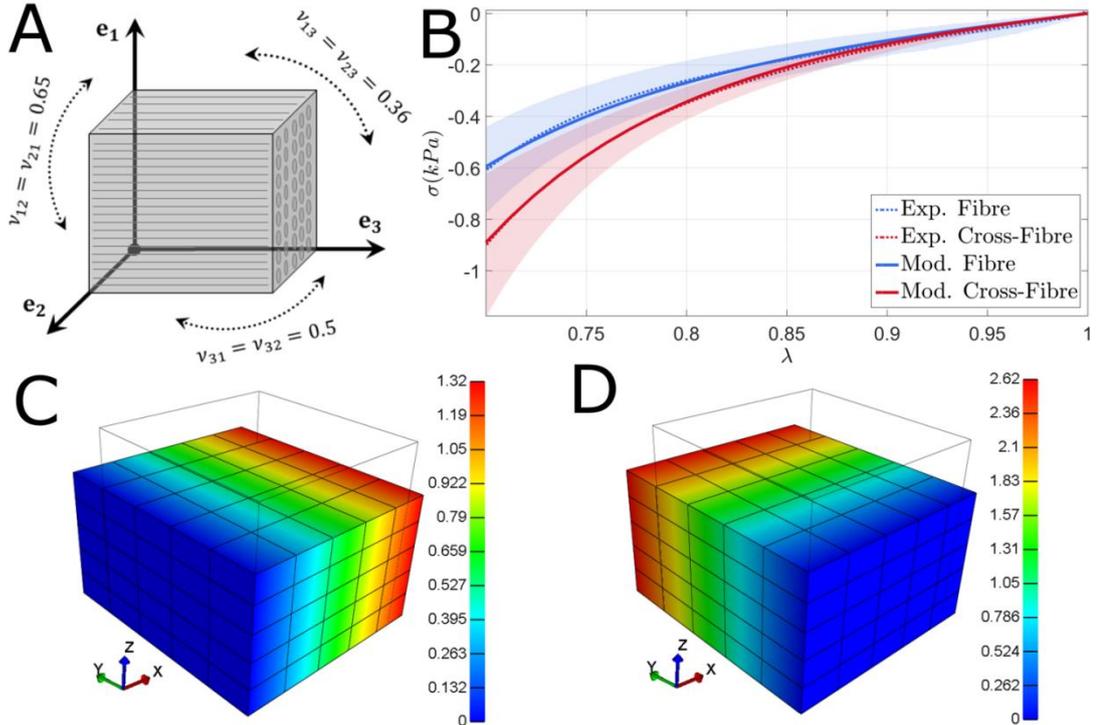

*Figure 5 A schematic of a muscle sample and Poisson's ratios with fibres along direction $e_3$ (A). The experimental (dashed) and model predicted (solid) stresses for fibre (blue), and cross fibre (red) loading respectively, transparent areas reflect standard deviations (B). The deformed finite element model showing predicted final displacements (units in mm) in the X (C) and Y direction (D).*

# 7. Summary and concluding remarks

This paper discusses Ogden hyperelastic formulations in terms of tension-compression asymmetry. Ogden hyperelasticity is presented in terms of Seth-Hill strains and it is shown how the properties of the Seth-Hill measures are reflected in the model behaviour. In addition, it is shown how the use of only positive or only negative Ogden coefficients may lead to a model where all behaviour is dictated by either tension or compressive processes respectively. In addition, these formulations exhibit a tension-compression asymmetry the degree of which cannot be separately controlled from the degree of non-linearity. A simple hybrid form is therefore proposed providing separate control over the tension-compression asymmetry. It is demonstrated how this form relates to a newly introduced generalised strain tensor class which includes both the tension-compression asymmetric Seth-Hill class and the tension-compression symmetric Bažant class. If the control parameter is set to $q = 0.5$, a tension-compression symmetric form involving Bažant strains is obtained with the property $\Psi(\lambda_1, \lambda_2, \lambda_3) = \Psi\left(\frac{1}{\lambda_1}, \frac{1}{\lambda_2}, \frac{1}{\lambda_3}\right)$. The symmetric form may be desirable for the definition of ground matrix contributions allowing all deviation from the symmetry to stem solely from fibrous reinforcement. Such an application is also presented demonstrating the use of the proposed formulation in the modelling of the non-linear elastic and transversely isotropic behaviour of skeletal muscle tissue in compression (the model implementation and fitting procedure have been made freely available).

Whether or not a material is tension-compression asymmetric should be demonstrable from multidirectional experiments. However, for biological tissues a limited set of experiments is often conducted, e.g. only indentation[34], or only tension[51], or only compression[43]. For ground matrices the exact nature, in terms of tension-compression symmetry or asymmetry, is unknown, and perhaps independent testing is not possible. Therefore, for these cases, the assumption of symmetry or asymmetry may be equally justified from this perspective. However, since the presented work allows for the independent control of the degree of asymmetry, and the formulation of a symmetric form if desired, it is more elegant than current modelling approaches whereby the tension-compression characteristics follow, somewhat arbitrarily, as a side-effect of a particular choice of constitutive parameters.

Special cases for models employing all positive or all negative Ogden coefficients are the Neo-Hookean models and first order Ogden formulations in general. Both are frequently used for modelling of biological tissues, the former is commonly used in ground matrix formulations[52] and the latter is popular for modelling of non-linear isotropic behaviour (e.g. skeletal muscle tissue[33] and skin[34]). However, it is here shown that a particular choice of Ogden coefficients not only affects the degree of non-linearity but also the degree of tension-compression asymmetry.

Although this paper treats Ogden formulations, the arguments can be extended to models incorporating singular terms involving $I_1(\mathbf{C})$ only. The degree of tension-compression asymmetry present in such formulations can be made controllable in a way equivalent to that presented here, simply by also including terms involving $I_1(\mathbf{C}^{-1})$.

The presented hyperelastic formulations may aid researchers in independently controlling the degree of tension-compression asymmetry from the degree of non-linearity, and in the case of anisotropic materials may aid in studying the role played by, either the ground matrix, or the fibrous reinforcing structures, in generating asymmetry.

## Acknowledgements

The authors would like to thank Dr. Alexander E. Ehret (ETH Zürich) for useful feedback on the initial draft of this manuscript.